\documentclass[pra, aps,twocolumn,showpacs,amsmath,amssymb,floatfix]{revtex4-2}
\usepackage{epsf}
\usepackage{graphicx}
\usepackage{epstopdf}
\usepackage{bm}
\usepackage{relsize}
\usepackage[normalem]{ulem}
\usepackage[usenames]{color}
\usepackage{float}
\usepackage{amsmath}
\usepackage[utf8]{inputenc}

\def \be {\begin{equation}}
\def \ee {\end{equation}}
\def \bea {\begin{align}}
\def \eea {\end{align}}
\def \p {\partial}
\def\bee{\begin{eqnarray}}
\def\eee{\end{eqnarray}}
\def \BC {\begin{cases}}
\def \EC {\end{cases}}
\def \m {\mathbf }

\begin{document}
	
	\title{
Beatings of ratchet current  magneto-oscillations in GaN-based grating gate structures: manifestation of spin-orbit band splitting. \\
	}
	
	\author{  P.~Sai$^{1,2,3},$  S.~O.~Potashin$^{4},$  M.~Szo{\l}a$^1,$  D.~Yavorskiy$^{1,5},$
 G.~Cywi\'nski$^{1,2},$  P.~Prystawko$^6,$ J.~{\L}usakowski$^5,$   S.~D.~Ganichev$^{1,7},$
    S. Rumyantsev$^{1},$ W.~Knap$^{1,2,8}$ and  V.~Yu. Kachorovskii$^{1,4}$}
	
\affiliation{\vspace{0.3cm}$^1$CENTERA Laboratories, Institute of High Pressure
Physics, Polish Academy of Sciences, 01-142 Warsaw, Poland} \affiliation{$^2$Centre for Advanced Materials and Technologies, Warsaw
University of Technology, 02-822 Warsaw, Poland	} \affiliation{$^3$V.
Ye. Lashkaryov Institute of Semiconductor Physics,  03680 Kyiv,
Ukraine} \affiliation{$^4$Ioffe Institute, 194021 St. Petersburg,
Russia} \affiliation{$^5$Faculty of Physics, University of Warsaw,
02-093 Warsaw, Poland}
\affiliation{$^6$Institute of High Pressure Physics, Polish Academy of Sciences, 01-142 Warsaw, Poland} \affiliation{$^7$Terahertz Center, University
of Regensburg, 93040 Regensburg, Germany}

\affiliation{$^8$Laboratoire Charles Coulomb, University of
Montpellier and Centre national de la recherche scientifique, 34950 Montpellier, France}
	
\begin{abstract}
We report on the study of the magnetic ratchet effect in AlGaN/GaN heterostructures superimposed with lateral superlattice formed by dual-grating gate structure. We demonstrate that irradiation of the superlattice with terahertz beam results in the \textit{dc} ratchet current, which shows giant magneto-oscillations in the regime of Shubnikov de Haas oscillations. The oscillations have the same period and are in phase with the resistivity oscillations. Remarkably, their amplitude is greatly enhanced as compared to the ratchet current at zero magnetic  field, and the envelope of these oscillations exhibits large beatings as a function of the magnetic field. We demonstrate that the beatings are caused by the spin-orbit splitting of the conduction band. We  develop a theory  which gives a good qualitative  explanation  of all  experimental observations and allows  us  to extract  the  spin-orbit splitting constant   $\alpha_{\rm SO}= 7.5 \pm 1.5$ meV\text{\AA}.  We also discuss how our results are modified by plasmonic effects   and  show that these effects become more pronounced with decreasing the  period of the gating gate structures down to  sub-microns.
\end{abstract}
	\date{\today, v.2}
	
	\maketitle
	\section{Introduction}

One of the most important tasks of modern optoelectronics is to	
provide efficient conversion of high-frequency terahertz (THz)
signals into a \textit{dc} electrical response, for reviews see, e.g.,
Refs.~\cite{Rev7,Rev6,Rev5,Rev3,Rev4,Rev2,Glazov2014,Koppens2014,Rev1}.
In the last decades, the focus
of research in this direction was on the
periodic structures like  field effect transistor (FET) arrays, grating gate, and multi-gate structures.
Such structures also 	
attract  growing  interest   as   simple examples of tunable
plasmonic 	crystals~\cite{Aizin1, Azin2, my1,  Azin3, Wang1}.
Plasmonic crystals already demonstrated excellent performance as  THz
detectors~\cite{allen1,allen2,allen4,Muravjov,allen6,Cai2014}, in close agreement with the numerical
simulations~\cite{32A,30A,31A,popov,popov2015}.  They are also
actively	studied as possible emitters or amplifiers  of THz
radiation~\cite{29A,new3,Tombet2020}. 	

Importantly, a 	non-zero {\it
dc} photoresponse requires some 	asymmetry of the system, which
would determine the direction of the 	produced {\it dc} current.
Generation of  \textit{dc} electric 	current in response to \textit{ac} electric field
in  systems with broken inversion 	symmetry is usually called the
ratchet effect which was studied both theoretically and 	
experimentally in a great number of systems, for reviews see, e.g.,
Refs.~\cite{rachet1,rachet4,28,32,33,34}. For effective radiation
conversion to \textit{dc} signal in periodic  structures, there
should be  strong  built-in  asymmetry inside the unit cell of the
plasmonic crystals. In particular, the ratchet {\it dc} current can be
induced by the electromagnetic wave incident on the spatially
modulated system provided that the wave 	amplitude  is  also
modulated but is 	phase-shifted   in space, for review see
Ref.~\cite{rachet4}.  On the theoretical level, the ratchet current arises already  in
non-interacting approximation (so called electronic ratchet).
 Electronic ratchets were discussed in the
two dimensional systems with lateral
gratings~\cite{rachet4,but13,but15,but14,Golub,budkin,Olbrich2016,24}
or arrays of asymmetric
dots/antidots~\cite{Chepelianskii2007,Sassine2008,Ermann2011,KannanPortal12}.
Electron-electron (ee) interaction can dramatically increase ratchet current due to plasmonic effects~\cite{Popov2011,Popov13,Otsuji2013,Watanabe2013,Hurita2014,Boubanga,Wang2015,Faltermeier2015,Rozhansky2015}. 	

Although the ratchet effect was treated theoretically and observed
experimentally in diverse low dimensional  spatially-modulated
structures, some basic issues of this effect still remain
puzzling. One of the interesting questions that has not yet been
discussed in the literature 	is the manifestation of the effects
of spin-orbit (SO) interaction in the ratchet effect. 	In this
paper, we  address this question. We study ratchet effect in magnetic
field (in what follows, we call it magnetic ratchet effect) in the
regime of Shubnikov de Haas (SdH) oscillations  and  demonstrate
that it is dramatically modified  by SO interaction. Specifically,
we report on the observation of the magnetic ratchet effect in the
lateral GaN-based superlattice formed by dual-grating gate (DGG)
structure. The specific property of the GaN systems as compared to
other 2D structures including graphene based structures is a very
high value of Rashba spin-orbit coupling (at least ten times larger
than in GaAs-based
structures)~\cite{Weber2005,21A,22A,23A,24A,AAA1,Wu,Dietl2014,25A}, which
is caused by high build-in electric field existing in such polar materials. This
is, therefore, very favorable for observations of SO-induced effects.
We demonstrate,  both experimentally and theoretically,  that in
quantized magnetic fields THz excitation results in the  giant
magnetooscillations of the  ratchet current coming from Landau
quantization, which, due to large SO  band splitting, are  strongly
modulated as a function of the magnetic field.
We  reach a good qualitative  agreement  between our experimental results and theory
(compare, respectively,  Fig.~\ref{Fig5e} and Fig.~\ref{Fig1} below).

Our results provide a novel method to study the band spin splitting. Currently the most
widely used techniques are direct measurements of magnetoresistance
in SdH oscillations regime~\cite{Bychkov84p78}, the weak anti-localization
experiments~\cite{Knap96}, optical methods~\cite{Dyakonovbook} and
photogalvanic studies~\cite{Ganichev2014}.  Since SdH oscillations
in magnetoresistance regime and ratchet current correlate, these two
measuring methods are complementary, which gives the opportunity to
double check the results.

The possibility to increase the ratchet effect in the magnetic field
deserves special attention. Therefore, we start the paper  with the
discussion of  the key points of the   magnetic ratchet effect, see
Sec.~\ref{state of the art}. The  rest of the paper is organized as
follows. In Sec.~\ref{experiment} we present the experimental results
on magnetic ratchet effects in GaN-based structures. In the following
Sec.~\ref{theory}  we present the theory and compare its results with
the experimental data.  Sec.~\ref{plasmonic} is devoted to the discussion
of the plasmonic effects. Finally, in Sec.~\ref{conclusion} we
summarize the results.

\section{Ratchet effect in magnetic fields: state of the art}
\label{state of the art}

Physics of the ratchet effect becomes much richer if one applies the
magnetic field. Magnetic ratchet effect, which is  in some
publications called magneto-photogalvanic effect, was widely studied
in different semiconductor systems. The magnetic ratchet effect can
be induced  even in the case of homogeneous graphene with structure
inversion asymmetry, see e.g.~\cite{Falko1989, Belkov2005,
Tarasenko2005,Tarasenko2011,Drexler2013}. The effect is sensitive to
disorder and can be tuned by the gate voltage~\cite{Drexler2013}.
Furthermore, the theoretical consideration of the magnetic ratchet effect
in graphene and bi-layer graphene showed that  it can be
substantially enhanced  under the 	cyclotron resonance (CR)
condition~\cite{Kheirabadi2018}.  Most recently it has been shown
theoretically and observed experimentally that magnetic ratchet
effect also can be drastically enhanced by 	deposition of asymmetric
lateral potential introduced by asymmetric 	periodic metallic
structure on top of a structure with two dimensional electron
gas~\cite{Budkin2016,golub,Faltermeier2018,Hubman2020}.  Remarkably,
magnetic ratchet strongly increases (by more than two orders  of
magnitude) in the SdH oscillations regime. Physically, this happens due to a very
fast oscillations of the resistivity with the Fermi energy, and
consequently, with the electron concentration. As a result, inhomogeneous (dynamical and static) density modulations induced by the electromagnetic wave leads to a very strong response.

Importantly, the responsivity in the regime of SdH oscillations
increases not only in grating gate structures but also in single
FETs~\cite{Sakowicz2008,Lifshits 2009,Boubanga2009,Klimenko2010}.
Although  the general physics of enhancement in both cases is
connected with fast oscillations  of resistivity, there is an
essential difference. In grating gate structures the shape of typical
\textit{dc} photoresponse roughly reproduces  resistance oscillations, while
in single FETs the typical response is $\pi/2$ shifted with respect
to resistance oscillations. The latter shift was explained
theoretically by hydrodynamic model in Ref.~\cite{Lifshits 2009} and demonstrated experimentally in Ref.~\cite{Klimenko2010}. The key idea is as follows. Transport scattering rate
$\gamma(n) $ in the SdH regime sharply depends on the dimensionless
electron concentration $n=(N-N_0)/N_0$ (here $N_0$ is background
concentration and $N$ is the concentration in the channel). Expanding $\gamma(n)\approx \gamma(0)~ +~ \gamma'(0)
n$ with respect to small $n,$ one finds that a nonlinear term,
$\gamma'(0) n \m v$,   appears in the Navier Stockes equation, where
$\m v$ is the drift velocity. This is sufficient to give a nonzero
response, which  in a single FET arises in the second order with
respect to external THz field  	(both $n$ and $\m v$ are linear with
respect to this field). Therefore, in this case, the response is
proportional 	to  first derivative of $\gamma' (0)$ with respect to
concentration (i.e. with respect to Fermi energy, $E_{\rm F}$), hence, it is
$\pi/2$ shifted in respect to the conductivity oscillations. By
contrast, in the  grating gate structures, the \textit{dc} response appears
only in the third order with respect to perturbation \cite{rachet4}.
As a consequence, the ratchet current is proportional to the second
derivative $\gamma''(0)$ (see discussion in Sec.~\ref{theory}) and therefore
roughly (up to a smooth envelope) reproduces resistance oscillations.

We will show that similar to other
structures~\cite{golub,Budkin2016,Hubman2020}, the amplitude of the
magnetooscillations is greatly enhanced as compared to the ratchet
effect at zero magnetic field. We experimentally demonstrate that the
photocurrent oscillates in phase with the longitudinal resistance
and, therefore, almost follows the SdH oscillations multiplied by a smooth
envelope. This envelope encodes information about cyclotron and
plasmonic resonances.   The most important experimental result is the
demonstration of the beatings of the  ratchet current oscillations. We
interpret these beatings assuming that they come from
SO splitting of the conduction band. The value of SO splitting
extracted from  the comparison of the experiment and theory is in
good agreement with independent measurements of SO band
splitting~\cite{Weber2005,21A,22A,23A,24A,AAA1,Wu,Dietl2014,25A}.

An important comment should  be made about the role of  the
ee interaction. Actually, the effect of the
interaction is twofold. First of all,  sufficiently fast ee
collisions drive the system  into the hydrodynamic regime.  We assume
that this is the case for our system and use the hydrodynamic  approach.
Secondly, ee-interaction   leads to plasmonic oscillations, so that a
new frequency scale, the plasma frequency, $\omega_p(q)$ appears in
the problem, where $q$ is the   inverse characteristic of the spatial
scale in the system. For a device with a short length, for example,
for a  single FET, $q$ is proportional to the inverse length of the
device. For periodic  grating gate structures, $q= 2\pi/L,$  where
$L$ is the period of the structure. At zero magnetic field, the  \textit{dc}
response    is dramatically enhanced in the  vicinity of plasmonic
resonance, $\omega=\omega(q)$  both for a single FET with asymmetric
boundary conditions~\cite{Dyakonov1993} and for periodic  asymmetric
grating gate structures~\cite{Rozhansky2015}. Also, the response essentially depends on
the polarization of the radiation.

Here, we calculate  analytically
the  \textit{dc} response in the quantizing magnetic field  within the hydrodynamic approximation for arbitrary
polarization of the radiation and analyzed  plasmonic
effects. One of our main prediction  is that for linearly polarized
radiation, the dependence of the ratchet current on the direction of
the polarization  appears only due to the plasmonic effects. We use
the derived expression to prove that for specific  parameters of our
structures the plasmonic effects are negligible, and as a
consequence, the \textit{dc} response does not depend on the polarization
direction. The latter issue is very important
for us since the direction of  linear polarization used in our
experiment was not well controlled.  We also argue
how to modify the structures in order to observe plasmonic resonances.

\section{Experiment}
\label{experiment}
\subsection{Experimental details}

We choose AlGaN/GaN heterostructure system for the experimental study
of the effect of SO splitting on magnetic ratchet effect. Important unique
properties of GaN system are the ability to form high density, high
mobility two dimensional electron gas (2DEG) on the AlGaN/GaN
interface, and large Rashba spin splitting of the conduction
band~\cite{Weber2005,21A,22A,23A,24A, AAA1, Wu,Dietl2014,25A}. Density of
2DEG and the band spin splitting in this system are about an order of
magnitude higher than that in AlGaAs/GaAs system. High carrier
density  is an important factor because as will be shown later the
amplitude of the photoresponse in the regime of the SdH oscillations
is proportional to the square of electron density.

\begin{figure}[ht]
	\centering
	\includegraphics[width=1\linewidth]{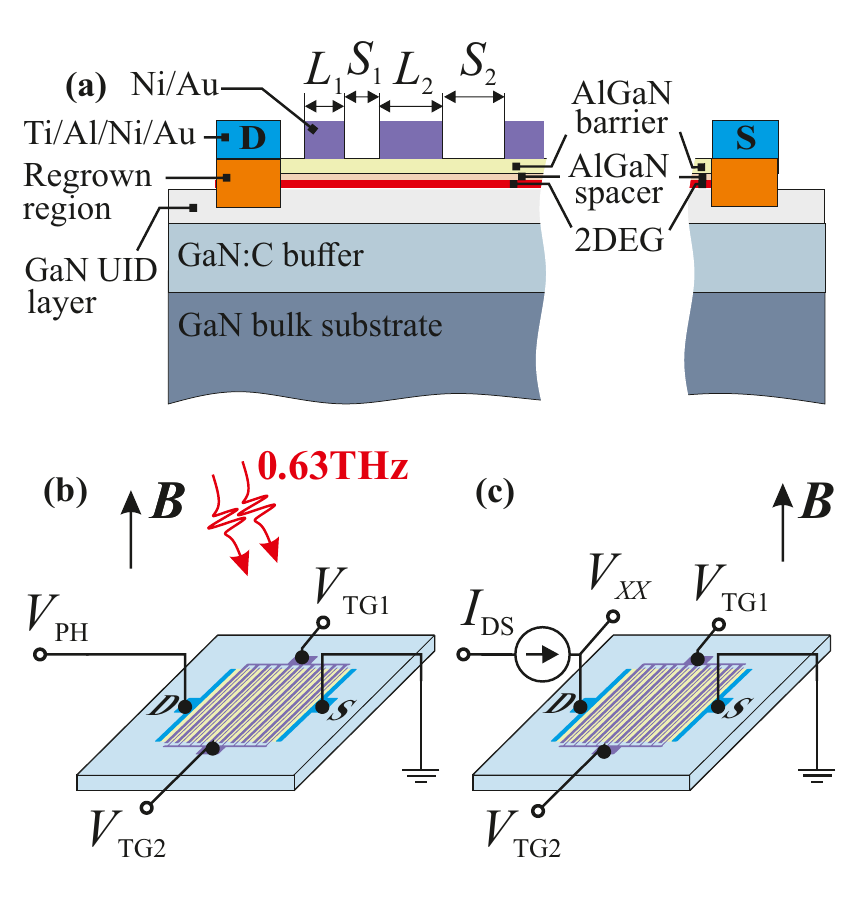}
	\caption{ (a) Cross sectional view of the sample heterostructure with DGG. (b) Scheme of the photoresponse measurements. (c) Scheme of the magnetoresistance measurements, $R_{xx} = V_{xx}/I_{\rm DS}$.
	}
	\label{Fig0e}
\end{figure}

AlGaN/GaN heterostructures were grown by Metalorganic Vapour Phase
Epitaxy (MOVPE) method in the closed coupled showerhead $3\times2$
inch Aixtron reactor  (Aixtron, Herzogenrath, Germany). The
epi-structure consisted of 25~nm $\rm{Al_{0.25}Ga_{0.75}N}$ barrier
layer, 1.5~nm $\rm{Al_{0.66}Ga_{0.37}N}$ spacer, 0.9~$\mu$m
unintentionally doped (UID) GaN layers, and 2~$\mu$m high resistive
GaN:C buffer, see Fig.~\ref{Fig0e}(a). Growth of all mentioned
epilayers was done on the bulk semi-insulating GaN substrates, grown
by the ammonothermal method~\cite{Dwilinski2008}. In this method high
resistivity of substrates (typically no less than $10^{9}
\Omega\cdot\rm{cm}$) was
obtained by compensation of residual oxygen, incorporated during
ammonothermal growth, by Mg shallow acceptors.

\begin{figure}[!h]
	\centering
	\includegraphics[width=1\linewidth]{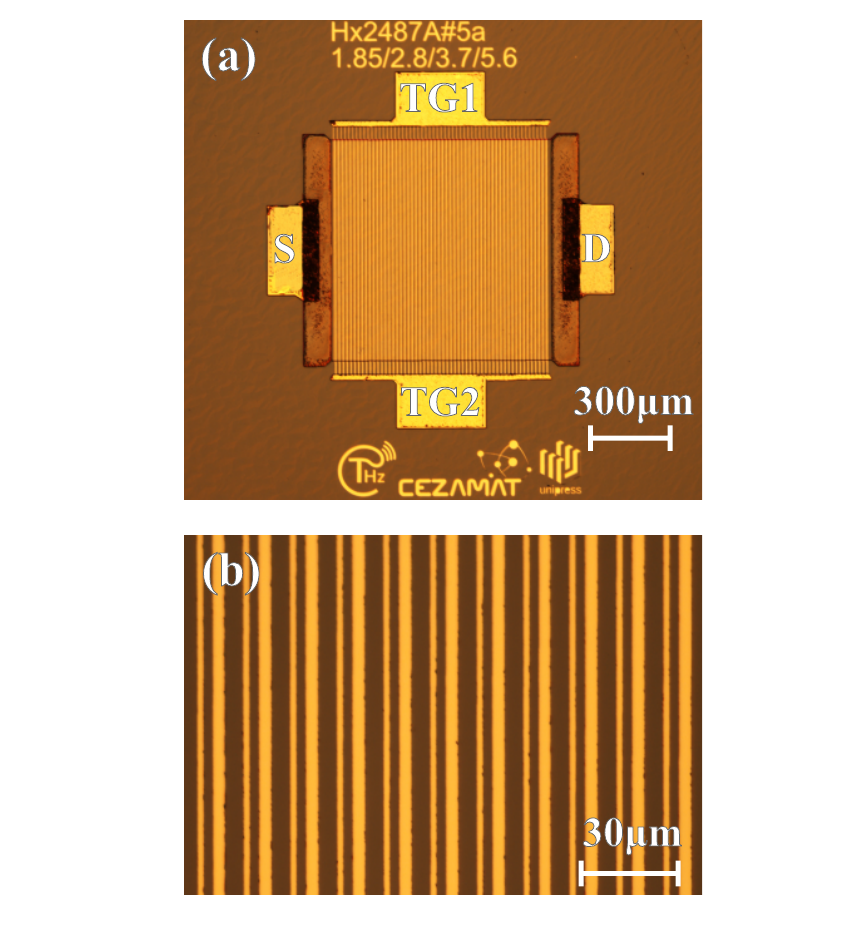}
	\caption{ Nomarski contrast microscope photos of investigated asymmetric DGG (a), where TG1 and TG2 – two multi-finger top gate electrodes, S and D – source and drain electrodes respectively; magnified active region of asymmetric DGG (b).}
	\label{Fig1e}
\end{figure}

The structures lithography processing was performed using a
commercial 405~nm laser writer system (Microtech, Palermo, Italy).
Devices were isolated from each other by shallow 150~nm mesas
etched by Inductively Coupled Plasma – Reactive Ion Etching (ICP-RIE)
(Oxford Instruments, Bristol, UK). In order to form the drain and
source ohmic contacts Ti/Al/Ni/Au (150/1000/400/500 {\AA}) stacks
were deposited on the MOVPE regrown heavily doped sub-contact regions
(for the detailed information of the regrowth technique see
Ref.~\cite{Wojtasiak2018}). Source and drain contacts, see
Fig.~\ref{Fig0e},  were annealed at 780 C$^\circ$ in a nitrogen
atmosphere for 60~s. This procedure yielded reproducible ohmic
contacts with resistances in the range of
$0.1$-$0.3\quad\Omega\cdot\rm{mm}$. Finishing fabrication step was the
deposition of Ni/Au (100/300 {\AA}) in order to form the DGG superlattice on the top of the AlGaN/GaN mesas. A schematic view
and Nomarski contrast microscope photos of fabricated devices are
shown in Figs.~\ref{Fig0e}(a) and~\ref{Fig1e}, respectively. The unit
cell of the DGG superlattice consisted of two gates of different
lengths ($L_1 = 1.85~\mu$m and $L_2 = 3.7~\mu$m) with different
spacings between them ($S_1 = 2.8\quad \mu$m and $S_2 = 5.6~\mu$m).
The cell was repeated 35~times resulting in a superlattice with a
total length of 500~$\mu$m.

\begin{figure}[t]
	\centering
	\includegraphics[width=1\linewidth]{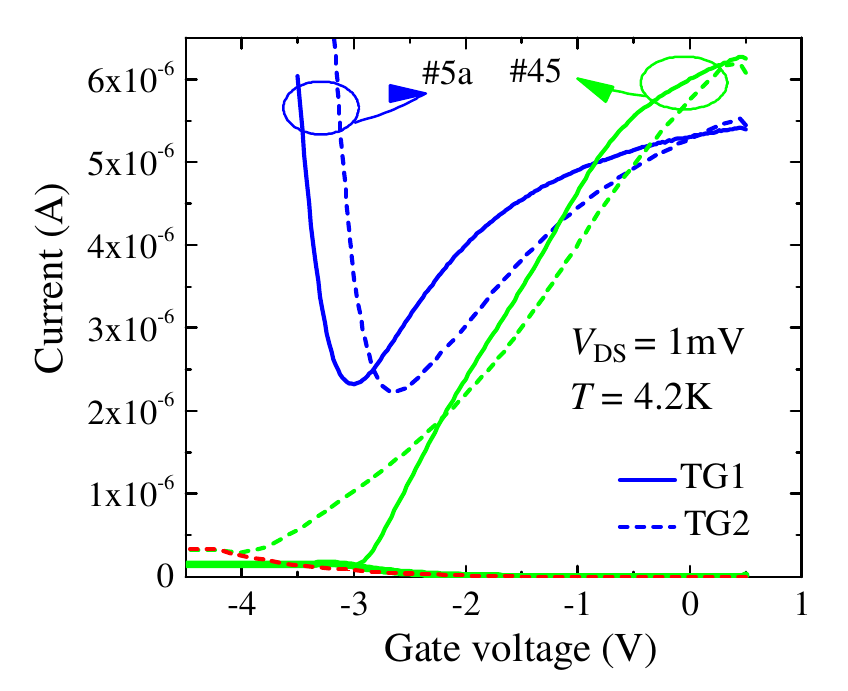}
	\caption{ Transfer current voltage characteristics for two representative devices. Red dashed line shows the gate leakage current for one of the devices.}
	\label{Fig2e}
\end{figure}

All wide gates were connected forming the multi-finger top gate
electrode TG1, see Fig.~\ref{Fig0e}(b,c) and Fig.~\ref{Fig1e}(a).
Similarly connected narrow gates formed the gate electrode TG2.
Independent bias voltages ($V_{\rm TG1}, V_{\rm TG2}$) could be
applied to wide and narrow gates.  The width of the whole structure
was 0.5~mm yielding the total active area $A = 0.25$~mm$^{2}$. The total gate area was~$\approx 0.1~\rm{mm^2}$. This
is a large area, which is $\sim$4 orders of magnitude bigger than that for
the ''standard'' transistor with gate length and width of 0.1~$\mu$m
and 100~$\mu$m, respectively. This made very challenging the
fabrication of the described DGG transistor with a reasonably small
gate leakage current. Figure~\ref{Fig2e} shows two examples of the
transfer current voltage characteristics of the studied devices.
Current in the subthreshold region is determined by the gate leakage
current (shown as a red dashed line for one of the devices). As seen,
the gate leakage current is rather small, significantly smaller than
the drain current even at a very low drain voltage of $V_{\rm
DS}=1$~mV. Even for those devices with relatively high gate leakage
current ($\#$5 in Fig.~\ref{Fig2e}) the drain current and, therefore
electron concentration can be changed several times by the gate
voltage.

Experimental set up is shown in Figs.~\ref{Fig0e}(b) and (c). As a
radiation source a frequency multiplier  from Virginia Diodes
Inc. with a radiation frequency of $f$=630 GHz was used to study the
ratchet effect. The radiation was guided onto the sample through a steel waveguide and was modulated at a frequency of about 173 Hz. External magnetic field up to 12 T was applied normally to 2DEG
plane, as shown in Fig. ~\ref{Fig0e}(b). The photoresponse, $V_{PH}$, was
measured in a cryostat at the temperature of 4.2 K in
the open circuit configuration using the standard lock-in technique.
Magnetoresistance was measured by applying a small $<1 \mu$A
current to the drain (see Fig.~\ref{Fig0e}c).

\begin{figure}[b]
	\centering
	\includegraphics[width=1\linewidth]{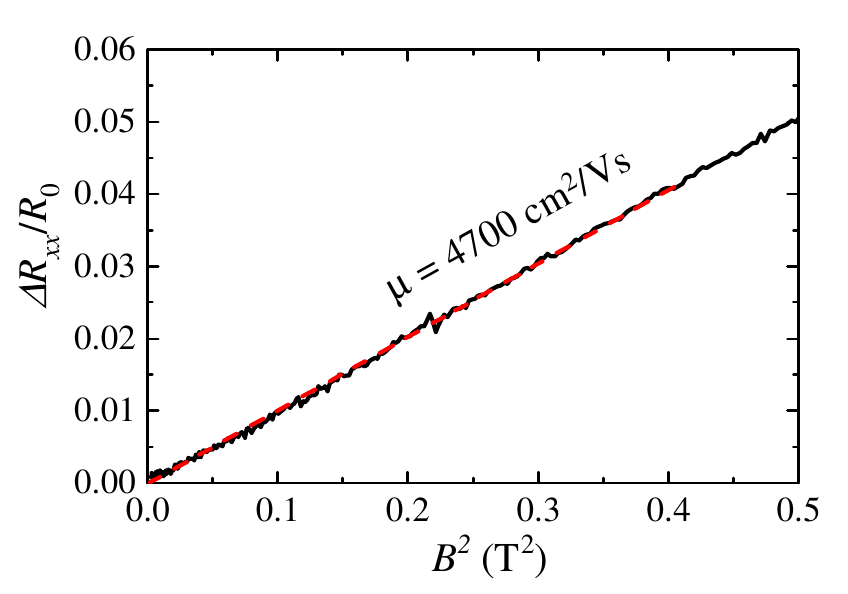}
	\caption{Magneto resistance as a function of $B^2$ in weak magnetic fields for a representative sample. Dashed line shows the linear fit.}
	\label{Fig3e}
\end{figure}

\subsection{Experimental results}

\begin{figure}[t]
	\centering
	\includegraphics[width=0.47\textwidth]{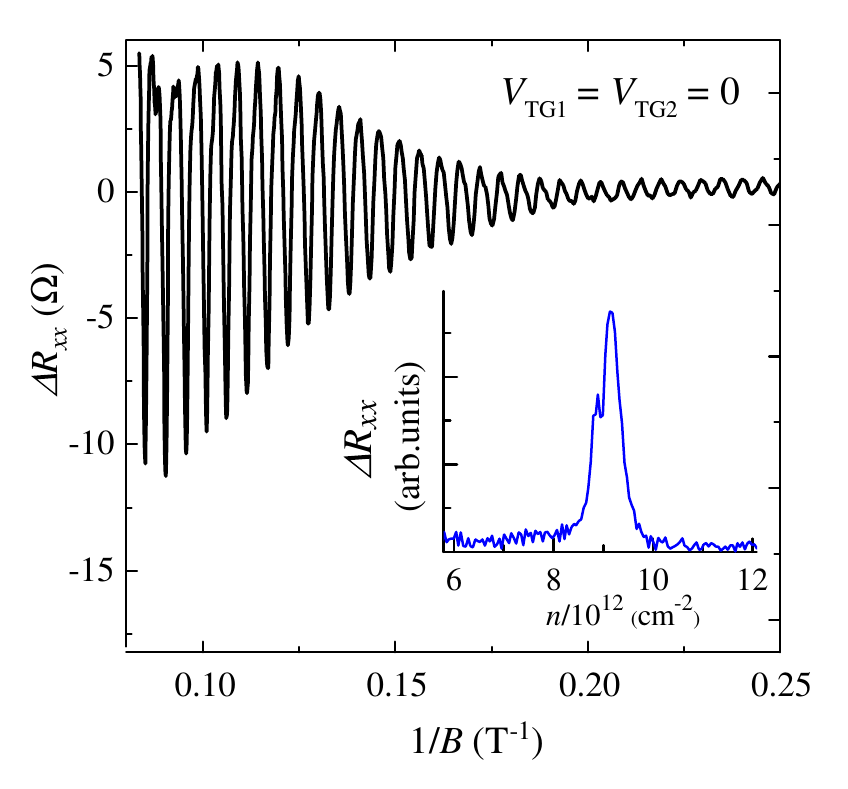}
	\caption{Resistance as a function of the inverse magnetic
field $1/B$ at  $V_{\rm TG1} = V_{\rm TG2} = 0$. The inset shows
shows the result of the Fourier transform of the oscillations in Fig.~\ref{Fig3e}
with frequency taken in the units of the electron concentration, $n=2e\nu/h$. }
	\label{Fig4e}
\end{figure}

First, we describe the results of the magnetotransport measurements, which are summarized  in Figs.~\ref{Fig3e}, and~\ref{Fig4e}. The overall shape of the DGG structures was close to the square. Therefore, in the magnetic field perpendicular to the drain to source plane investigated structures exhibited geometrical magnetoresistance~\cite{Lippman}. The full geometrical magnetoresistance is observed either in the disk Corbino geometry or in the  samples with $W>>L$.  For the arbitrary shaped rectangular samples, the geometrical magneto resistance can be approximated as~\cite{Schroder}:
\begin{align}\label{Exper}
	\frac{\bigtriangleup R}{R_0}\cong\left(\mu B\right)^2\left(1-0.54\frac{L}{W}\right).
\end{align}
This allowed us to extract the electron mobility. Figure~\ref{Fig3e} shows the experimental dependence of the magnetoresistance obtained for a weak magnetic field for one of the samples as a function of $B^2$. The estimate yields $\mu = 4700$~cm$^{2}$/Vs.

The concentration in the channel can be extracted from the magneto
resistance in the higher magnetic fields. Figure~\ref{Fig4e} shows
the resistance SdH oscillations as a function of the inverse magnetic
field $1/B$ measured for both top gate voltages equal to zero. The
concentration is given by
\begin{align}\label{Exper}
	N=\frac{2 e}{h\bigtriangleup (1/B)}=\frac{2 e \nu}{h},
\end{align}
where $\bigtriangleup(1/B)$ and $\nu$ are the period and frequency of SdH oscillations, respectively. The
 inset in Fig.~\ref{Fig4e} shows the result of the Fourier transform
 of the resistance magnetooscillations  with the frequency taken in
 the units of the electron concentration, $N=2e \nu/h$. Two peaks in the
 inset correspond to the concentrations $N=9.14 \times
 10^{12}$~cm$^{-2}$ and $N = 8.89 \times 10^{12}$~cm$^{-2}$, which we
 interpret as electron densities in the gate free and under the gate
 areas, respectively.

\begin{figure}[b]
	\centering
	\includegraphics[width=1\linewidth]{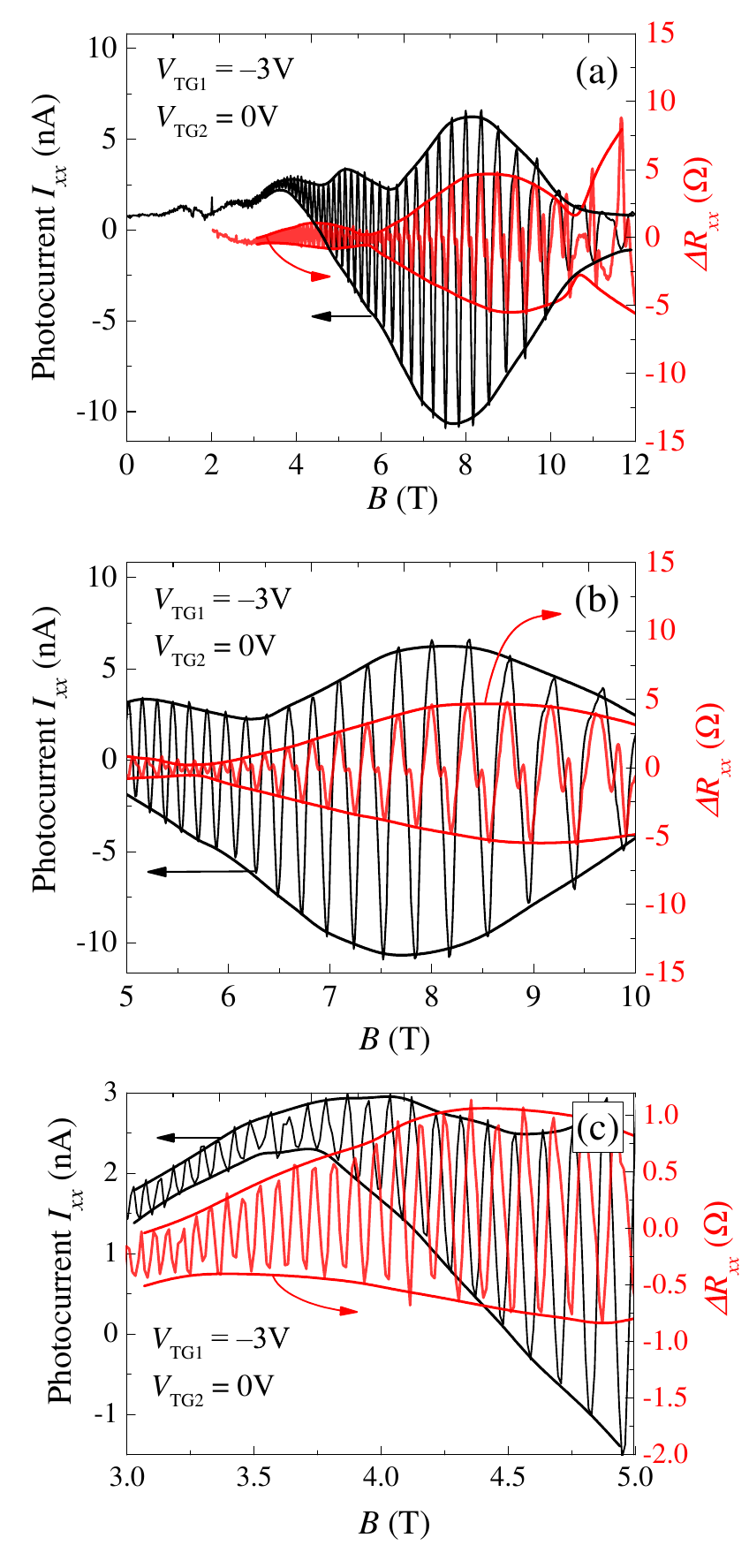}
	\caption{Photocurrent and resistance SdH oscillations as a function of magnetic field from $B=0$ to $B=12$~T (a). (b) and (c) show close look to intermediate and low magnetic field regions, respectively.}
	\label{Fig5e}
\end{figure}

Irradiating the unbiased structures we detected a photosignal caused
by the generation of the ratchet photocurrent.  Figure~\ref{Fig5e}(a)
shows the photoresponse measured for asymmetric gate voltages
applied: $V_{\rm TG1}=-3V, V_{\rm TG2}=0$. The photoresponse current
was calculated as $I_{xx}=V_{\rm PH}/R_{xx}$. In low and zero
magnetic fields the response is positive and weakly depends on the
magnetic field.  When gate voltages were changed to $V_{\rm TG1}=0,
V_{\rm TG2}=-3V$ the magnitude of the response was approximately  the
same, but of the negative sign. The change of the signal sign upon
inversion of the lateral asymmetry is a clear indication that the
observed photocurrent is caused by the ratchet effect, for review see
Ref.~\cite{rachet4}.  Indeed, the direction of the current is
controlled by the lateral asymmetry parameter~\cite{rachet4}
\begin{equation}
\label{Xi}
\Xi = \overline{|\bm E(x)|^2 \frac{dV(x)}{dx}},
\end{equation}
where $V
(x)$ and  $E(x)$ are spatially modulated by grating gates
static potential and electric field amplitude (over-line shows average over the modulation period).
Exchange of the gate voltages applied to
the TG1 and TG2 results in the change of sign of $dV/dx$ and,
consequently in the sign inversion of the ratchet current.

The increase of magnetic field results in the sign-alternating oscillations with the amplitude by orders of amplitude larger than the signal obtained for zero magnetic field. Moreover, the envelope of the oscillations exhibits large beatings as a function of magnetic field.  Comparison of the observed oscillations with the SdH magnetoresistance oscillations demonstrates that at high magnetic fields both, photocurrent and resistivity oscillations, have the same period and phase. Importantly, SdH effect also show similar beatings of the envelope function.  To facilitate the comparison of the phase of the oscillations, we zoom the data of the panel (a) for the range of fields $B= 6 \div 10 $~T in the panel (b) Fig.~\ref{Fig5e}.

The overall behavior of the observed current, besides beatings,
corresponds to that of the magneto-ratchet current most recently
detected in CdTe-based quantum wells~\cite{Budkin2016,golub} and
graphene~\cite{Hubman2020}. Importantly, the oscillations of the
magneto-ratchet current are in phase with the SdH oscillations. As we
discussed in Sec.~\ref{state of the art}, this differs from the
photocurrent magnetooscillations detected in single
transistors~\cite{Klimenko2010} described by the theoretical model of
Lifshits and Dyakonov~\cite{Lifshits 2009}.
We will show in Sec.~\ref{theory} that the
theory of the magneto-ratchet effect predicts  the photoresponse to be
 proportional to the second derivative of the magnetoresistance, and, hence describes well the experimental findings.
Note, that in the low magnetic field range,
see Fig.~\ref{Fig5e}(c), the oscillations of the photoresponse and
magnetoresistance are phase shifted. This is an indication that at
low magnetic fields we have at least two different competing
mechanisms of the detection. Below we focus on the case of
sufficiently high magnetic fields  and postpone discussion of
possible  competing mechanisms at low fields for future research.
Importantly, the theory  demonstrates that, in agreement with the experiment
(see Fig.~\ref{Fig5e}) the photoresponse in the regime of
SdH oscillations is significantly enhanced as compared to that at
zero magnetic field. This  point is very important in view of
possible applications and deserves  a special comment. The
theoretical limit for the response of the detectors based on the
direct rectification is defined by the device built-in nonlinearity.
In Schottky diodes and FETs the maximum current responsivity is approximately
$e/2\eta kT$, where $e$ is the elemental charge and $\eta$ is the
ideality factor of the Schottky barrier or subthreshold slope of FET
transfer characteristic~\cite{Cowley1966,Kachorovskii2013}. The
responsivity of the real device is usually orders of magnitude smaller
due to the parasitic elements and not perfect coupling. However,
increasing of the theoretical limit still should be beneficial for
the increasing of the responsivity in real devices.
Reducing temperature indeed leads to the responsivity increase but
only to a certain limit. As shown in~\cite{Klimenko2012} the
temperature decrease below 30K does not lead to the increase of
responsivity. This low-temperature saturation is caused by the
increase of factor $\eta$ with temperature decrease which is known in
Schottky diodes and FETs~\cite{Sze}. In the magnetic field under the
regime of SdH oscillations resistance of FET very sharply depends on
the gate voltage and provides the opportunity to go beyond $e/2\eta
kT$ limit. We can speculate that about an order of magnitude increase
of the responsivity in an external magnetic field in Fig.~\ref{Fig5e}
demonstrates the increase of the physical responsivity beyond the
fundamental limit.

Importantly, the theory presented below also describes well the observed oscillations of the envelope amplitude of the photoresponse, which are clearly seen in Fig.~\ref{Fig5e}. It shows that the oscillations are due to the spin-splitting of the conduction band and can be used to extract this important parameter.

\section{Theory }
\label{theory}

Above, we experimentally demonstrated that the photocurrent oscillates and the shape of these oscillations almost follow the SdH resistivity oscillations. In this section, we consider gated 2DEG and demonstrate that the results can be well explained within the hydrodynamic approach.

The effect which we discuss here is present for the system with an arbitrary energy spectrum. However,
calculations are dramatically simplified for the parabolic spectrum, so
that we limit calculations to this case only. We assume that electron
density in the structure is periodically modulated by built-in static
potential  and   study optical \textit{dc} response to linearly-polarized
electromagnetic radiation  which is also spatially modulated with the
phase shift $\varphi$ with respect to  modulation of the  static
potential.

A variation of individual gate voltages of the DGG lateral structure allows one to change  controllably the sign of $V(x)$ and, consequently,  the direction of the ratchet current. Furthermore,  the phase of the oscillations of the magnetic ratchet current is  sensitive to the orientation of the radiation electric field vector with respect to the DGG structure as well as to the radiation helicity. In the latter case, switching from right- to left-circularly polarization results in the phase shift by $\pi$, i.e., at constant magnetic field the helicity-dependent contribution to the current changes the sign. We consider magnetic field induced modification of the zero $B$-field electronic ratchet effect. We also analyze our results theoretically for different relation between $\omega$  and $\omega_p(q)$ having in mind to find signature of the plasmonic effects.

\subsection{Model}

We model     electric field of  the radiation, $\m E(x,t)= \m E(x)e^{-i \omega t}+c.c., $  and the  static potential, $V,$  as follows
\begin{align}\label{Ex}
	E_x (x,t) & = [1+ h \cos(q x+\varphi)]  E_0 \cos \alpha  \cos \omega t ,
	\\
	\label{Ey}
	E_{y}(x,t) & = [1+ h \cos(q x+\varphi)]  E_0 \sin \alpha  \cos (\omega t+\theta),\\
	\label{V}
	V(x)  & = V_0\cos qx,
\end{align}
where $h \ll 1$ is the modulation depth, $\varphi$ is the phase, which determines the asymmetry of the modulation, $\alpha$ and $\theta$ are constant phases describing the  polarization of the radiation. These phases are connected with the standard Stockes parameters (normalized by $E_0^2$) as follows:
\be
\begin{aligned}
	& P_0=1, ~P_{\rm L1}=\sin(2 \alpha) \cos \theta,
	\\
	& P_{\rm L2}=\cos(2\alpha),~P_{\rm C}=\sin(2\alpha) \sin \theta.
	\label{Stockes}
\end{aligned}
\ee
Within this model, asymmetry parameter [see Eq.~\eqref{Xi}] becomes
\begin{equation}
\label{Xi-exact}
	\Xi= \frac{E_0^2 h V_0 q \sin \varphi  }{4} .
\end{equation}
As seen, $\Xi$ is proportional to the sine of the spatial phase shift $\varphi.$

Hydrodynamic   equations  for concentration  and velocity   look
\begin{align}
	\label{n}
	& \frac{\p n}{ \p t}+ {\rm div}~[(1+n) \m v]=0, \\
	\label{v}
	& \frac{\p \m v}{ \p t}+ (\m v {\nabla}) \m v+  \gamma (n) \m v
	+\boldsymbol{ \omega}_c\times \m v + s^2 {\nabla} n =\m a.
\end{align}
Here
\be n= \frac{N -N_0}{N_0}, \ee
$N=N(x,t)$  is the concentration in the channel and $N_0$ its equilibrium value,
\be \m a= - \frac{e \m E}{m} + \frac{e}{m}  {\nabla} V,\quad \m E=
\left[
\begin{array}{c}
	E_x \\
	E_y \\
\end{array}
\right], \ee
$\boldsymbol{\omega}_c= e \m B/m_{\rm eff} c$  is the cyclotron frequency in the external  magnetic field $\m B$, $s$ is the  plasma waves velocity, $\gamma( n)=1/\tau_{\rm tr}(n)$ is momentum relaxation rate. The non-linearity is encoded in hydrodynamic terms $\p (n \m v)/\p x,$ $(\m v \boldsymbol{\nabla})\m v  $ as well as independence of transport scattering rate  on the  concentration.  Specifically, we use the approach suggested in Ref.~\cite{Lifshits 2009}.  We assume that $\gamma(n)$ is controlled by the local value of the electron concentration, $n,$ which, in turn, is  determined  by the local value of the Fermi energy, $n(\m r)   =   (E_{\rm F} (\m r) - E_{\rm F}^0)/E_{\rm F}^0$   (here we   took into account that the 2D density of states is energy-independent). Due to the SdH oscillations, scattering rate is an oscillating function of $E_{\rm F},$ and, consequently, oscillates with $n.$ In the absence of SO coupling, $\gamma(x,t)= \gamma[n(x,t)] $ is given by \cite{Lifshits 2009}
\be \gamma(x,t)= \gamma \left\{\!  1-  \delta \cos\!\!
\left[\frac{2\pi E_{\rm F} (x,t)}{\hbar \omega_c}\right]\!\right\},
\label{shd} \ee
where
\be \delta =\frac{ 4\chi } {\sinh \chi} \exp\!\!\left(\!-
\frac{\pi}{\omega_c\tau_q} \!\right) \ee
is the amplitude of the SdH oscillations, $$\chi= \chi(\omega_{\rm c})= \frac{2\pi^2 T}{\hbar \omega_{\rm c}},$$ $T$ is the temperature in the energy units, $E_{\rm F}(x,t)=E_{\rm F}[1+n(x,t)]$ is the local Fermi energy, which is related to concentration in the channel as $N(x,t)=\nu E_F (x,t)$ (here $\nu$ is the density of states), and $\tau_q$ is quantum scattering time, which can be strongly renormalized by electron-electron collisions in the hydrodynamic  regime. We assume that   $ 2\pi^2 T + \pi \hbar /\tau_q  \gg  \hbar \omega_{\rm c}.$   Then,
\be \delta \ll 1, \label{ineq} \ee
and the  second term in the curly brackets in Eq.~\eqref{shd} is very small.  Hence $\gamma(n)$ is very close to the value of transport scattering rate  $\gamma$ at zero magnetic field.

Eq.~\eqref{shd} can be generalized for  the case of non-zero SO coupling by using results of Refs.~\cite{golub,tarasenko,averkiev}
\begin{align}
	&\gamma(n)=\gamma \left[\!  1-  \frac{ 4\chi}{\sinh \chi} \right.
	\label{shd-SO0}
	\\
	\nonumber
	& \left. \times   \exp\!\!\left(\!- \frac{\pi}{\omega_{\rm c}\tau_{\rm q}}
	\!\right) \cos\!\! \left(\frac{2\pi E_{\rm F}(1+ n)}{\hbar \omega_c}\right)
	\cos\left( \frac{2 \pi \Delta  }{\hbar \omega_{\rm c}}\right)\!\right].
\end{align}
Here we assumed that there is linear-in-momentum    spin-orbit splitting of the spectrum, $E(k) = \hbar^2 k^2/2 m  \pm  \Delta,$ where
\be \Delta= \alpha_{\rm SO} k_{\rm F}, \label{Delta} \ee
is characterized by coupling   constant $\alpha_{\rm { SO}}.$ Experimentally measured values of $\alpha_{\rm SO}$ lays between 4-10 meV\text{\AA} \cite{Weber2005,21A,22A,23A,24A, AAA1, Wu,Dietl2014,25A}. For such values of $\alpha_{\rm SO}$  and typical values of the concentration, one can assume  $\Delta \ll E_{\rm F}$ and neglect dependence  of $k_{\rm F}$ on $n$. Equation \eqref{shd-SO0} was derived under the assumption that the quantum scattering time $\tau_{\rm q}$ is the same in two spin-orbit split subbands.  Actually,  this assumption is correct only for the model of  short range  scattering potential where both transport and quantum scattering rates  are momentum independent.  For any finite-range potential, the quantum scattering times in two subbands differ,  because of small difference of the Fermi wavevectors $k_1=\sqrt{2m (E_{\rm F} +\Delta)  }/\hbar$ and $k_2=\sqrt{2m (E_{\rm F} -\Delta)  }/\hbar.$   Denoting these times as $\tau_1$ and $\tau_2,$ we get instead of Eq.~\eqref{shd-SO0}:
\be \frac{\gamma(n)}{\gamma}=   1-  \frac{ 2\chi}{\sinh \chi}\! \sum
\limits_{i=1,2}\!\! \exp\!\!\left(\!- \frac{\pi}{\omega_{\rm
c}\tau_{\rm i}} \!\right) \cos\!\! \left[\!\frac{2\pi E_{i}(n)}{\hbar
\omega_c} \!\right],\! \label{shd-SO} \ee
where
\be   E_1(n)=E_{\rm F} (1+n) +\Delta,\quad E_2(n)=E_{\rm F} (1+n)
-\Delta. \label{Ei} \ee

Detailed microscopical calculation of $\tau_{1,2}$ for  specific model of the scattering potential is out of the scope of the current work. Here, we use $\tau_{1,2}$ as fitting parameters.

Let us now  expand  $\gamma(n)$ near the Fermi level:
\be \gamma(n) = \gamma(0) + \gamma'(0) n  + \gamma''(0)
\frac{n^2}{2}, \label{gamma-n} \ee
where  $\gamma'$   and   $\gamma''$ are, respectively, first and second derivatives  with respect to $n$ taken  at the Fermi level. Since  oscillations are very fast, we assume
\be \frac{\gamma'}{\gamma} \propto \frac{\gamma''}{\gamma'} \propto
\frac{ E_{\rm F}}{\hbar \omega_{\rm c}} \gg 1. \label{condition} \ee
Due to these inequalities oscillating contribution to the ratchet current can be very large and substantially exceed zero-field value \cite{Hubman2020}.

Here, we focus on SdH oscillations of the ratchet current, so that we only keep oscillations  related to dependence of $\gamma$ on $n$ and, moreover, skip  in Eq.~\eqref{gamma-n} the term  proportional  to $\gamma'.$

We use  the  same method of calculation  as one developed in Ref.~\cite{Rozhansky2015}. Specifically,  similar to impurity-dominated regime \cite{rachet4} we  use the perturbative expansion of $n$ and $\m v$  and  \textit{dc}  current,
\be \m J_{dc}= -eN_0
\left \langle (1+n)\m v \right \rangle_{t,x} \ee
over $E_0$ and $V.$ Non-zero contribution, $\propto E_0^2 V_0$, arises  in the order $(2,1)$  [see Eq.~\eqref{Xi}].

\subsection{Calculations and results }

Let us formulate the key steps of calculations. Due to the  large
parameter, ${ E_F}/{\hbar \omega_c} \gg 1,  $   the main contribution
to the  rectified ratchet current comes from the non-linear term
$\gamma'' \mathbf v  {n^2}/{2}$ in the r.h.s. of Eq.~\eqref{v} [see
also Eq.~\eqref{gamma-n}].  We, therefore, neglect all other nonlinear terms in the hydrodynamic
equations. Calculating $n$ and $\m v$ in
linear (with respect to $E_0$ and $V$) approximation, substituting
the result into non-linear term and averaging over time and
coordinate, we get $\gamma''(0)\langle  \mathbf v
{n^2}\rangle_{x,t}/2 \neq 0. $  Next, one can find rectified current
$\m J_{\rm dc}$  by   averaging of Eq.~\eqref{v} over $t$ and $x$.
This procedure is quite standard, so that we delegate it to the
Supplemental Material (similar calculations were performed in
Ref.~\cite{Rozhansky2015} for zero magnetic field). The result reads
\be \frac{\m J_{\rm dc}}{ J_0}=\frac{\gamma''(0)}{\gamma} ~ \m R.
\label{Jstart} \ee
Here
\be J_0=-\left(\frac{eE_0}{2 m}\right)^2 \left(\frac{eV_0 q}{2 ms^2}\right) \frac{e N_0 h\sin\varphi} {\gamma^3} \ee
is the frequency and magnetic field independent  parameter  with dimension of the current (physically, $J_0$ gives the typical value of current for the case,when all frequencies are of the same order, $\omega \sim \omega_c \sim q s \sim  \gamma \sim 1/\tau_{1,2}$), the dimensionless factor ${\gamma''}/{\gamma}$ accounts for SdH oscillations and dimensionless vector
\be \m R = \frac{ \gamma^4
	(P_0\hspace{0.2mm} \m a_0+ P_{\rm L1}\hspace{0.2mm} \m a_{\rm L1}
	+ P_{\rm L2}\hspace{0.2mm} \m a_{\rm L2}+ P_{\rm C} \hspace{0.2mm}\m a_{\rm C}) }
{\left|\omega_{\rm c}^2 - (\omega-i \gamma)^2\right|^2
	(\gamma^2+\omega^2)(\gamma^2+\omega_{\rm c}^2)|D_{\omega q}|^2  }.
\label{R} \ee
depends on the radiation polarization encoded in the vectors $$\m a_i = \left[
\begin{array}{c}
a_{ix} \\
a_{iy} \\
\end{array}
\right]$$ ($i=0,{\rm L1, L2,C} $)  and  also contains  information about cyclotron  and magnetoplasmon  resonances which occur for $\omega = \omega_{\rm c}$ and $\omega = \sqrt{\omega_{\rm c}^2+s^2q^2},$ respectively. The latter resonance appears due to the factor $D_{\omega q}$ in the denominator of Eq.~\eqref{R}. Analytical expressions for $\m a_i$ and $D_{\omega q}$ are quite cumbersome and presented in the Supplementary material [see Eqs.~\eqref{a0}, \eqref{aL1}, \eqref{aL2}, \eqref{aC} and \eqref{D}].

The second derivative of the scattering rate with respect to $n$ is calculated by using Eq.~\eqref{shd-SO}:
\begin{align}
	&g(\omega_{\rm c})= \frac{\gamma''(0)}{\gamma} =
	\frac{ 2 \chi(\omega_{\rm c})}{\sinh[ \chi (\omega_{\rm c})]}
	\left ( \frac{2\pi E_{\rm F}}{\hbar \omega_c}\right)^2
	\label{gamma''0}
	\\
	\times
	&
	\sum \limits_{i=1,2}\!\! e^{\displaystyle - {\pi}/{\omega_{\rm c}\tau_{\rm i}}
	} \cos  \left[{2\pi E_{i}(0)}/{\hbar \omega_c} \right].
	\nonumber
\end{align}
Here, $E_1(0)=E_{\rm F}+\Delta, ~E_2(0)=E_{\rm F}-\Delta $ [see Eq.~\eqref{Ei}]. The function $g(\omega_{\rm c})$ rapidly oscillates due to the factors $ \cos \left[{2\pi E_{i}(0)}/{\hbar \omega_c} \right].$   For $\omega_c\to 0$ this function goes to zero due to the Dingle factors $\exp( - {\pi}/\omega_{\rm c}\tau_{\rm i}),$ so that discussed mechanism has nothing to do with the zero field ratchet effect.

The  smooth envelope of  the function $g(\omega_{\rm c})$ reads
\begin{align}
	&\tilde g(\omega_{\rm c})=  \frac{ 2 \chi(\omega_{\rm c})}{\sinh[ \chi (\omega_{\rm c})]}
	\left ( \frac{2\pi E_{\rm F}}{\hbar \omega_c}\right)^2
	\label{tilde-g}
	\\
	\times
	&
	\left| e^{\displaystyle - {\pi}/{\omega_{\rm c}\tau_{\rm 1} + 2 \pi i \Delta/\hbar \omega_{\rm c}  }}
	~+~ e^{\displaystyle - {\pi}/{\omega_{\rm c}\tau_{\rm 2} - 2 \pi i \Delta/\hbar \omega_{\rm c}  }}\right|.
	\nonumber
\end{align}
Function $g(\omega_{\rm c})$ shows rapid SdH oscillations with the beats due to the spin orbit coupling. As seen from the behavior of the envelope function $\tilde g(\omega_{\rm c}),$ the beats are most pronounced for $\tau_1=\tau_2,$ when $\tilde g(\omega_{\rm c}) $ is proportional to $ \cos(2\pi\Delta/\hbar \omega_{\rm c})$ and therefore vanishes at the values of $\omega_{\rm c}^n,$ obeying $2\pi \Delta/\omega_{\rm c}^n=\pi/2+ \pi n.$   For $\tau_1\neq \tau_2, $ envelope function is nonzero at these points, $\tilde g(\omega_{\rm c}^n)\neq 0 ,$ and beats are less pronounced.

Now, we are ready to explain why  the response in the  SdH oscillation regime is much larger than   at zero magnetic field. The enhancement of the response as compared to the case $B=0,$ is due to the  factor
\be  \delta \left(\frac{2 \pi E_{\rm F}}{\hbar \omega_{\rm c}}
\right)^2 \gg  1.  \label{ineq1} \ee
One can check that for experimental values of parameters inequalities
Eq.~\eqref{ineq} and \eqref{ineq1} are satisfied simultaneously in a
wide interval of magnetic fields, $  1 < B < 7 $ T. It is also
important that due to the  coefficient  $E_{\rm F}^2$ in the
$g(\omega_{\rm c})$ the response increases with the concentration in
contrast to conventional transistor operating at $B=0,$ where
response is inversely proportional to the concentration at high
concentration  \cite{Dyakonov1993} and saturates at low concentration
when a transistor is driven below  the threshold \cite{Knap2002}. This means that the use of AlGaN/GaN system  for detectors operating
in the SdH oscillation regime is very advantageous because of the extremely high concentration of 2DEG.

\begin{figure}[t]
	\centering
	\includegraphics[width=1\linewidth]{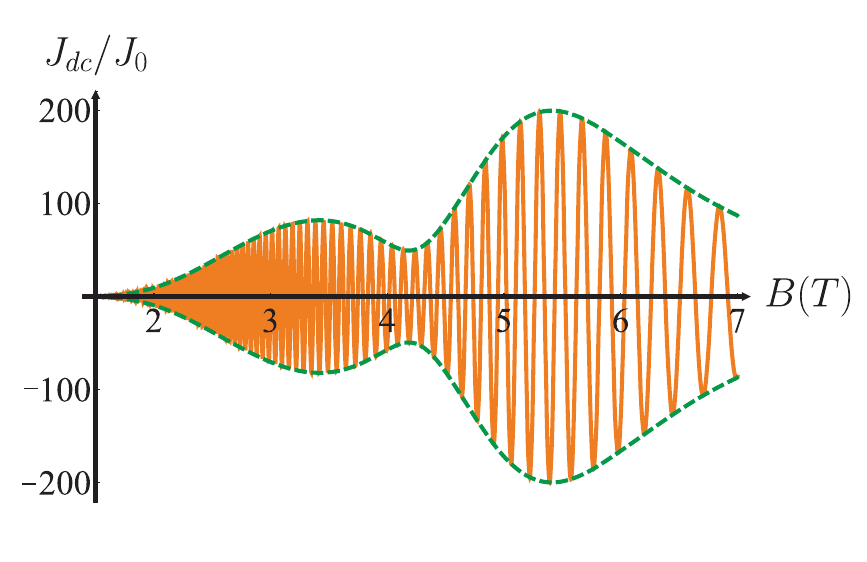}
	\caption{Theoretically calculated
ratchet magnetooscillations for the following parameters: 	
$\epsilon=9,\alpha_{\rm SO}= 7.5$ meV\text{\AA}, $T=4$K, 		
$m_{\rm eff}=0.23 m_{\rm e}, d=2.5\cdot 10^{-6}$cm, $L=15 \cdot 10^{-4}$cm,
$N_{\rm 0}=8 \cdot 10^{12}$cm$^{-2},~ \alpha=0,\tau_{\rm tr}= 1.2{\times 10^{-12}{\rm s}},~ \tau_{1} =	1.4
{\times 10^{-12}{\rm s}},~ \tau_{2}=10^{-12}$~s.
}
	\label{Fig1}
\end{figure}

Let us discuss the polarization dependence of the response. Importantly,
vectors $\m a_i,$  responsible for polarization dependence, contain
$q$-independent terms and terms proportional to $\omega_q^2=s^2q^2.$
The latter   describe  plasmonic effects. As seen from
Eqs.~\eqref{a0},\eqref{aL1}, \eqref{aL2},  for  small $q$ (or/and
small $s$),  vectors  $\m a_{\rm L1}$ and $\m a_{L2}$ are small,
$\propto q^2.$ In other words, for  our case  of linearly-polarized
radiation   with polarization directed by angle $\alpha,$ the
dependence  of the rectified current on $\alpha$ appears only due to
the plasmonic effects. For experimental values of the parameters, the
value of plasmonic frequency, $s q$ was sufficiently small, about
$0.7 \cdot 10^{12}$ s$^{-1},$ which is  much smaller than the
radiation frequency (for $f=0.6 $ THz we get $\omega=2\pi f \approx
3.8 \cdot 10^{12} $ s$^{-1}$).  As follows from  this estimate, the
plasmonic effects are actually small and can be neglected.

Then,   the response does not  actually depend on
polarization angle $\alpha.$  This justifies our experimental
approach, where $\alpha$ is not well controlled.  Within this
approximation, one can put $q\to 0 $ in Eqs.~\eqref{a0},\eqref{aL1},
\eqref{aL2}, and \eqref{D}. Then, the analytical expression for current
simplifies. In the absence of the circular component of  polarization
($P_{\rm C}=0$), we get

\be\label{Jxy}
   \left [\begin{array}{c}
           J_{\rm dc}^x
           \\
           J_{\rm dc}^y
         \end{array}\right]=
         \frac{2 J_0 ~g(\omega_{\rm c})~ \gamma^4 \omega_{\rm c}}
         {(\gamma^2+ \omega_{\rm c}^2)|(\omega+i \gamma)^2-\omega^2_{\rm c}|^2}\left [\begin{array}{c}
           -\omega_{\rm c}
           \\
           \gamma
         \end{array}\right].
   \ee
This expression simplifies even further in the resonant regime, $\omega \approx \omega_{\rm c} \gg \gamma$:

\be\label{Jxy-res}
  \left [\begin{array}{c}
           J_{\rm dc}^x
           \\
           J_{\rm dc}^y
         \end{array}\right]=
         \frac{  J_0 ~g(\omega_{\rm c})~ \gamma^4 }
         {2 \omega_{\rm c}^3[(\omega-\omega_{\rm c})^2 +\gamma^2]}\left [\begin{array}{c}
           -\omega_{\rm c}
           \\
           \gamma
         \end{array}\right].
   \ee This expression shows rapid oscillations, described by
function $g(\omega_{\rm c}),$ which envelope represent a   sharp
CR with the width $\gamma.$

\begin{figure}[t]
	\centering
	\includegraphics[width=1\linewidth]{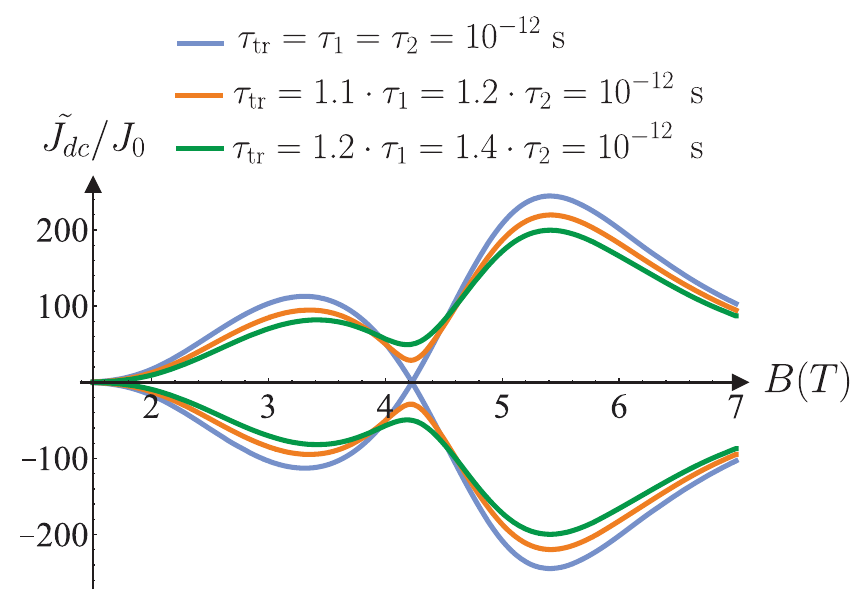}
	\caption{ Envelope of the ratchet current magnetooscillations  for different  ratio 		
of quantum times $\tau_1$ and $\tau_2$ (other parameters are the same as in Fig.~\ref{Fig1}).
}

	\label{Fig2}
\end{figure}

\begin{figure}[b]
	\centering
	\includegraphics[width=1\linewidth]{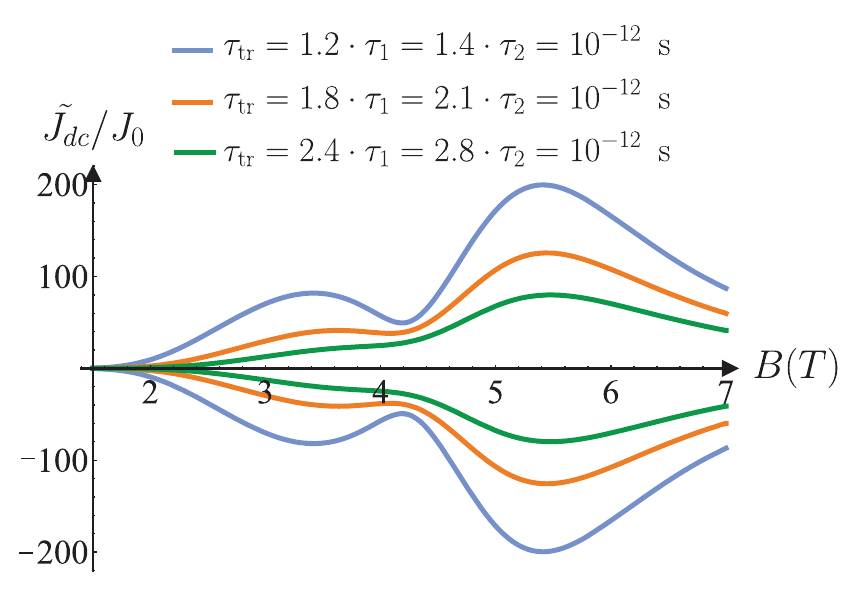}
	\caption{
	Envelope of the ratchet current magnetooscillations  for different values of the 	
	quantum times $\tau_1$ and $\tau_2$ with the fixed ratio $\tau_1/\tau_2$
(other parameters are the same as in Fig.~\ref{Fig1}).
	}
	\label{Fig4}
\end{figure}

\begin{figure}[t]
	\centering
	\includegraphics[width=1\linewidth]{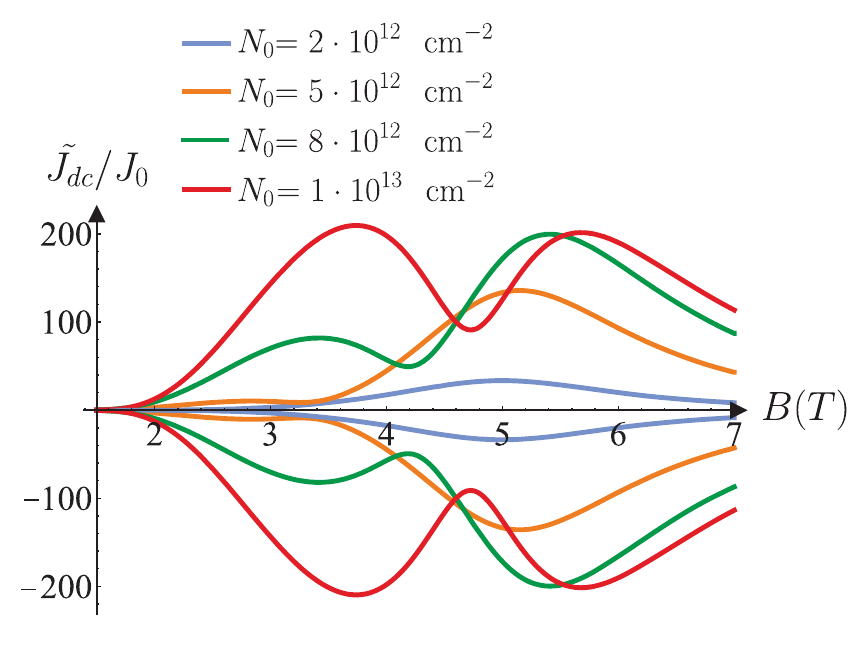}
	\caption{
	Envelope of the ratchet current magnetooscillations for  different 		values of the electron concentration
		(other parameters are the same as in Fig.~\ref{Fig1}).
}
	\label{Fig3}
\end{figure}

Let us now compare theoretical results with experimental observations.
In Fig.~\ref{Fig1} we plot the $x-$component of the  rectified
\textit{dc} current (this component was actually measured in the
experiment), calculated with the use of Eq.~\eqref{Jstart}, as a
function of the magnetic field. We assumed that the   radiation is
linearly-polarized along $x$ axis [$\alpha=\theta=0$, $P_{\rm
C}=P_{\rm L1}=0,~P_{\rm L2}=1,$ see Eqs.~\eqref{Ex},\eqref{Ey},  and
\eqref{Stockes}]   and  used  experimental values of  parameters:
$m_{\rm eff}=0.23 m_{\rm e},  d=2.5\cdot 10^{-6}$cm, $L=15 \cdot
10^{-4}$cm, $\epsilon=9,$ $T=4$K, $n=8 \cdot
10^{12}$cm$^{-2},~\tau_{\rm tr}= \gamma^{-1}=10^{-12}$s,
$\omega=3.8\cdot 10^{12}$s$^{-1}$. The best fit was obtained for $\alpha_{\rm SO}= 7.5 \pm 1.5$ meV\text{\AA}, in accordance with previous measurements of SO band splitting \cite{Weber2005,21A,22A,23A,24A, AAA1, Wu,Dietl2014,25A}.  We used $\tau_{1,2}$ as the fitting parameters
choosing $\tau_{\rm tr}= 1.2 ~ \tau_{1} = 1.4 ~ \tau_{2}.$  We see
that exactly this   behavior  is observed in the  experiment  (see
Fig.~\ref{Fig5e}).  Most importantly, we reproduce experimentally
observed  beats of SdH oscillations using the value of $\alpha_{\rm
SO} $  consistent with previous experiments. In Figs.~\ref{Fig2},
\ref{Fig3}, and \ref{Fig4} we show dependence of the smooth envelope
of the current, $\tilde J_x,$  on magnetic field for different values
of $\tau_{1,2}$ and different concentrations. As we explained above,
the most pronounced modulation is obtained for $\tau_1=\tau_2$ (see
Fig.~\ref{Fig2}). Dependence on concentration appears  both due to
the factor $(E_{\rm F}/\hbar \omega_{\rm c})^2$ in $g(\omega_{\rm
c})$ and due to the  dependence of $\Delta $ on $k_{\rm F}.$

\begin{figure}[!h]
	\centering
	\includegraphics[width=1\linewidth]{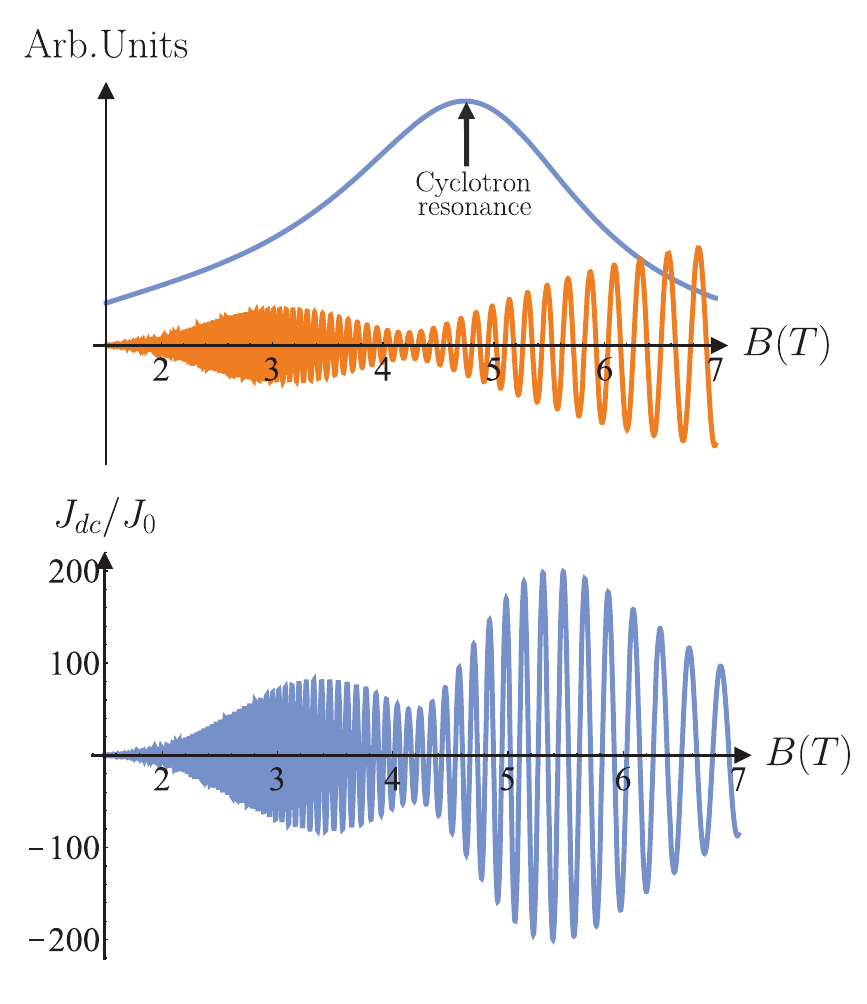}
	\caption{
		DC response for small  $q$ [shown in bottom panel,
described by Eq.~\eqref{Jxy}] is given by the product of smooth function $R_x$
[blue curve in upper panel], which shows CR	and rapidly oscillating function $g(\omega_{\rm c})$
[orange curve in upper panel, described by Eq.~\eqref{gamma''0}],
which contains beats of SdH oscillations.
	}
	\label{Fig5}
\end{figure}

\begin{figure}[!h]
	\centering
	\includegraphics[width=1\linewidth]{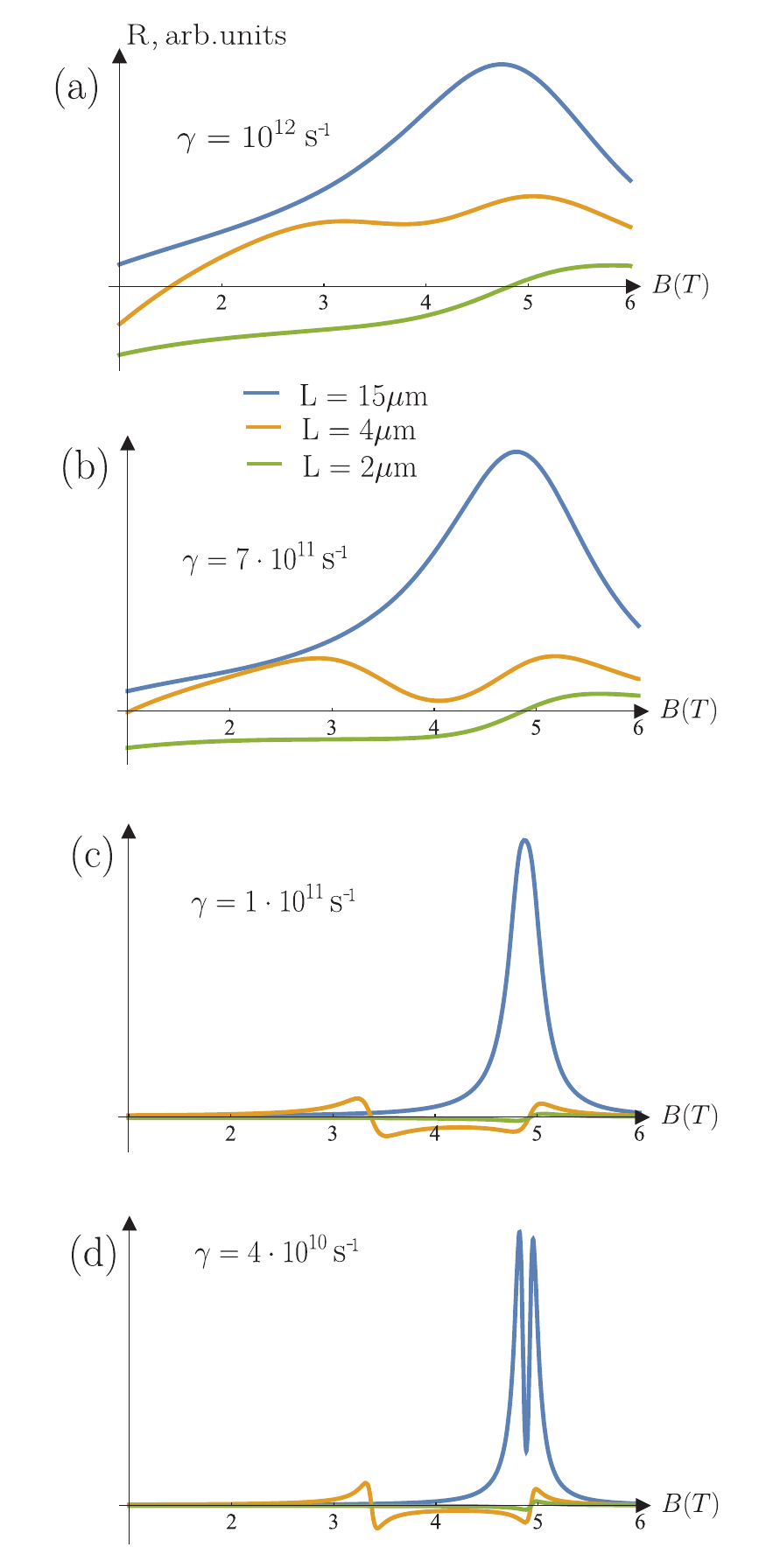}
	\caption{
Cyclotron and magnetoplasmon resonances in the smooth function $R_x$ [see Eq.~\eqref{R}]
for different values of  wavevector $q=2\pi/L$  (determined by size of the	unit cell $L$)
and transport scattering rate $\gamma.$  For large values of  $\gamma$ (a)	
plasmonic effects are fully  negligible   for 		small $q$ (large $L$) and function
$R_x$ shows  only CR at $\omega_{\rm c}=\omega$ (this case corresponds to our experimental situation).
		With decreasing $L$ there appears a week  plasmonic resonance at the value of $\omega_{\rm c}$
  given by $\sqrt{\omega^2-s^2q^2}.$ 		For smaller $\gamma$ (b,c), both cyclotron and
  magnetoplasmonic resonances become sharper. For very small 		$\gamma$ (d),
  plasmonic resonance appears even for very small $q$.
	}
	\label{Fig6}
\end{figure}

Evidently, Eq. \eqref{Jxy} can be presented as a product  of a smooth function
describing  cyclotron resonance (CR) and rapidly oscillating function, which encodes information
about SO splitting. This is illustrated in Fig.~\ref{Fig5}.

\subsection{Role of the  plasmonic  effects}\label{plasmonic}

Above, we demonstrated that  plasmonic effects can be neglected for
our experimental parameters and, as a consequence, the response is
insensitive to the direction of the linear polarization. However, the
role of plasmonic effects is not fully understood. The point is that
the existing ratchet theory assumes a weak coupling with a
diffraction grating. In such a situation, the plasmon wave vector,
which determines  plasma oscillations  frequency, is set by the total
lattice period: $q = 2 \pi / L$. In the experiment,  $L$ = 13.95 $\mu$m,
i.e. is very large and, as a consequence, the plasma frequency
corresponding to the full period is small. This frequency does not
appear in the experiment, as follows from the theoretical  pictures
presented above (see Fig.~\ref{Fig6}a). If we assume, that
the coupling is not so weak, then the plasmons  determined only
by the gate region, should show up.
Then $q$ is determined only by the gate length $L_{\rm g}<L$ and it should manifest
itself, as it is shown in Fig.~\ref{Fig6}. Namely,  plasma wave effects should lead
to the plasmonic  splitting of the CR.

The role of the plasmonic effects can be enhanced either by decreasing the period of the structure, which implies increasing of $q$  or  by decreasing   transport scattering rate $\gamma$ as illustrated in Fig.~\ref{Fig6}, where cyclotron and magnetoplasmon resonances in the smooth function $R_x$ [see Eq.~\eqref{R}] are shown for different values of   $L$   and  $\gamma.$   Although, for large values of $\gamma$ (Fig.~\ref{Fig6}a) plasmonic effects are fully  negligible for small $q$ (large $L$) and function  $R_x$ shows  only CR at $\omega_{\rm c}=\omega,$  with decreasing $L$ there appears a week  plasmonic resonance at $\omega_{\rm c}=\sqrt{\omega^2-s^2q^2}.$ For smaller $\gamma$ (Fig.~\ref{Fig6} b, c), both cyclotron and magnetoplasmonic resonances become sharper. For very small $\gamma$ (Fig.~\ref{Fig6}c), plasmonic resonance appears even for very small $q.$

\section{Conclusion}\label{conclusion}

To conclude, we presented  observation of the magnetic ratchet effect
in GaN-based structure superimposed with lateral superlattice formed
by dual-grating  gate (DGG) structure. We showed that  THz excitation
results in the  giant magnetooscillation of the  ratchet current in
the regime of SdH oscillations. The amplitude of the oscillations is
greatly enhanced as compared to the ratchet effect at zero magnetic
field. We  demonstrate that the photocurrent oscillates as the second
derivative of the longitudinal resistance and, therefore, almost
follow the SdH resistivity oscillations multiplied by a smooth
envelope. This envelope encodes information about cyclotron
resonance.   One of the most important experimental results is the
demonstration of beats of the ratchet current oscillations.  We
interpret these beats theoretically assuming that they come from SO
splitting of the conduction band. The value of SO splitting extracted
from  the comparison of the experiment and theory is in good agreement
with independent measurements of SO band splitting. We also discuss conditions required for the observation of magnetoplasmon resonances.

\section{Acknowledgements}

A financial support of the IRAP Programme of the Foundation for Polish Science (grant MAB/2018/9, project CENTERA) is greatfully acknowledge. The partial support by the National Science Centre, Poland allocated on the basis of Grant Nos. 2016/22/E/ST7/00526, 2019/35/N/ST7/00203 and NCBIR WPC/20/DefeGaN/2018 is acknowledged. S.D.G. thanks DFG-RFBR project (numbers Ga501/18-1 and 21-52-12015) and  Volkswagen Stiftung Cooperation Program (97738). The work of S.P. and V.K. was  funded by RFBR, project numbers 20-02-00490 and 21-52-12015,  and   by  Foundation for the Advancement  of Theoretical Physics and Mathematics.

\begin{widetext}
	\section{ Supplemental Material }
	\label{Supplemental}
	{\it In this Supplemental Material to the paper ``Beatings of ratchet current  magneto-oscillations in GaN-based grating gate structures: manifestation of spin-orbit band splitting.'' we present technical details of the calculations}
	$$$$
	We start with continuity equation and Navie-Stokes equation written in   components (notations are the same as in the main text)
	\begin{equation}
	\begin{array}{l}
	\begin{split}
	&\frac{\p n}{ \p t}+ \frac{\p v_{x}}{\p x}=0, \\
	&\frac{\p v_{x}}{ \p t}+\gamma v_{x}+\omega_{c} v_{y}+s^{2}\frac{\p n}{\p x}=\frac{e}{m}\left\{V_{0}\frac{\p \cos{q x}}{\p x}-E_{0x}(1+h_{x}\cos{(q x +\phi)})\cos{\omega t}\right\}-\frac{1}{2}\gamma''n^{2}v_{x},  \\
	&\frac{\p v_{y}}{ \p t}+\gamma v_{y}-\omega_{c} v_{x}=-\frac{e E_{0y}}{m}\left(1+h_{y}\cos{(q x +\phi)}\right)\cos{\left(\omega t+\theta\right)}-\frac{1}{2}\gamma''n^{2}v_{y}.
	\end{split}
	\end{array}
	\end{equation}
	
	Here
	\begin{equation}
	E_{0x}=E_{0}\cos{\alpha},E_{0y}=E_{0}\sin{\alpha}
	\end{equation}
	describe  linearly-polarized wave.
	Following procedure described in the main text,  we take into account in these equations only non-linearity related to dependence of $\gamma$ on $n$ (assuming that term with second derivation of $\gamma''(0)$ dominates) linearizing all other terms.
	
	We search the solution  as perturbative  expansion over $E_0$ and $V_0.$
	The   nonzero rectified electric current
	\begin{equation}
	\mathbf{J}_{dc}=-eN_{0}\mathbf{j}_{dc},    \mathbf{j}_{dc}=\left\langle\left(1+n\right)\mathbf{v}\right\rangle_{t,x}
	\end{equation}
	appears in the third order  ($\propto E_0^2 V_0$): $\mathbf{j}_{dc}\approx\mathbf{j}_{dc}^{2,1},$.
	In order to find  $\mathbf{j}_{dc}^{2,1},$ we need to calculate
	\begin{equation}
	\begin{split}
	-\frac{\gamma''}{2}\left\langle n^2\mathbf{v}\right\rangle_{x,t}=-\frac{\gamma''}{2}\left\langle n^{0,1}(x) n^{1,0}(x,t)\mathbf{v}^{1,0}(x,t)\right\rangle_{x,t}\varpropto E^{2}V_{0}\sin{\phi}\ne 0.
	\end{split}
	\end{equation}
	Hence, we need to calculate  $n^{0,1}(x),~ n^{1,0}(x,t),~ \mathbf{v}^{1,0}(x,t)$.
	\subsection{Calculation in the order (0,1)}
	In this order, one puts $E_0=0,$ and there is only correction to the electron concentration induced by the inhomogeneous static potential
	\begin{equation}
	\mathbf{v}^{0,1}=0, \quad n^{0,1}(x)=\left(\frac{eV_0}{ms^2}\right)\cos{qx}.
	\end{equation}
	\subsection{Calculation in the order  (1,0)}
	We search for solution in the form
	\begin{equation}
	\begin{split}
	&\mathbf{v}^{1,0}(x,t)=\left(\frac{eE_0}{m}\right)\left[\mathbf{\underline{V^{1,0}}}(t)+h\mathbf{\underline{\underline{V^{1,0}}}}(t)\sin{(qx+\phi)}\right],\\
	& n^{1,0}(x)=\left(\frac{eE_{0}h}{m}\right)\underline{\underline{N^{1,0}}}(t)\sin{(qx+\phi)}.
	\end{split}
	\label{notations}
	\end{equation}
	The current is expressed in terms of these notations as follows
	\begin{equation}
	\begin{split}
	\mathbf{j}_{dc}^{2,1}=-\frac{\gamma''}{2}\frac{eV_{0}N \sin{\phi}}{ms^{2}}\frac{1}{\gamma^2+\omega_{c}^2}\begin{pmatrix}
	\gamma& -\omega_{c}\\
	\omega_{c}& \gamma
	\end{pmatrix}\left\langle  \underline{\underline{N^{1,0}(t)}}\cdot \underline{\mathbf{V}^{1,0}(t)}\right\rangle_{t}.
	\end{split}
	\end{equation}
	The amplitudes  entering Eq.~\eqref{notations} can be found by substituting Eq.~\eqref{notations} in the starting equations and separating terms of the order  $(1,0)$
	
	\begin{equation}
	\begin{split}
	&\left(\gamma-i\omega\right)\underline{V_{x}^{10,\omega}}+\omega_{c}\underline{V_{y}^{10,\omega}}=-\frac{eE_{0x}}{2m}e^{-i\omega t}, \\
	&\left(\gamma-i\omega\right)\underline{V_{y}^{10,\omega}}-\omega_{c}\underline{V_{x}^{10,\omega}}=-\frac{eE_{0y}}{2m}e^{-i\omega t}e^{-i\theta}, \\
	\end{split}
	\end{equation}
	
	\begin{equation}
	\begin{split}
	&\left(\gamma-i\omega+i\frac{s^2 q^2 }{\omega}\right)\underline{\underline{V_{x}^{10,\omega}}}+\omega_{c}\underline{\underline{V_{y}^{10,\omega}}}=-\frac{eE_{0x}h_{x}}{2m}e^{-i\omega t} ,\\
	&\left(\gamma-i\omega\right)\underline{\underline{V_{y}^{10,\omega}}}-\omega_{c}\underline{\underline{V_{x}^{10,\omega}}}=-\frac{eE_{0y}h_{y}}{2m}e^{-i\omega t}e^{-i\theta}, \\
	&-i\omega \underline{\underline{N^{10}}}-q\underline{\underline{V_{x}^{10,\omega}}}=0.
	\end{split}
	\end{equation}
	Solution of these linear equations reads
	\begin{equation}
	\begin{pmatrix}
	\underline{V_{x}^{1,0}}& \\
	\underline{V_{y}^{1,0}}
	\end{pmatrix}=\frac{eE_{0}}{2m}\frac{e^{-i\omega t}}{(\omega+i\gamma)^2-\omega_{c}^{2}}\begin{pmatrix}
	-i(\omega+i\gamma)\cos{(\alpha)}-\omega_{c}\sin{(\alpha)}e^{-i\theta}& \\
	\omega_{c}\cos{(\alpha)}-i(\omega+i\gamma)\sin{(\alpha)}e^{-i\theta}
	\end{pmatrix}+c.c,
	\end{equation}
	
	\begin{equation}
	\begin{split}
	\begin{pmatrix}
	\underline{\underline{V_{x}^{1,0}}}& \\
	\underline{\underline{V_{y}^{1,0}}}
	\end{pmatrix}=-\frac{eE_{0}h}{2m}\frac{e^{-i\omega t}}{(\omega+i\gamma)D_{q\omega}}\begin{pmatrix}
	i\omega(\omega+i\gamma)\cos{(\alpha)}+\omega\omega_{c}\sin{(\alpha)}e^{-i\theta}& \\
	-\omega\omega_{c}\cos{(\alpha)}+i(\omega(\omega+i\gamma)-s^2q^2)\sin{(\alpha)}e^{-i\theta}
	\end{pmatrix}+c.c,
	\end{split}
	\end{equation}
	\begin{equation}
	\underline{\underline{N^{1,0}}}=q\frac{eE_{0}h}{2m}\frac{\left[(\omega+i\gamma)\cos{(\alpha)-i\omega_{c}\sin{(\alpha)}e^{-i\theta}}\right] e^{-i\omega t}}{(\omega+i\gamma)D_{q\omega}}+c.c.
	\end{equation}
	Here
	\begin{equation}
	D_{q\omega}=\omega(\omega+i\gamma)-\omega_{c}^{2}\frac{\omega}{\omega+i\gamma}-q^2s^2.
	\end{equation}
	Then, we find
	\begin{equation}
	\begin{split}
	&\left\langle  \underline{\underline{N^{1,0}(t)}}\cdot\underline{\mathbf{V}^{1,0}(t)}\right\rangle_{t}=\left(\frac{eE_{0}}{2m}\right)^2\frac{q h}{(\omega+i\gamma)D_{q\omega}\left[(\omega-i\gamma^2)-\omega_{c}\right]}\begin{pmatrix}
	C_x& \\
	C_y
	\end{pmatrix}+c.c,
	\end{split}
	\end{equation}
	where
	\begin{equation}
	\begin{split}
	&C_x=i[\cos^2{\alpha}~(\omega^2+\gamma^2)+\omega_{c}^2\sin^2{\alpha}-2\cos{\alpha}
	\sin{\alpha}(\omega_{c}\omega\sin{\theta}+\gamma\omega_{c}\cos{\theta})],\\
	&C_y=i[\cos^2{\alpha}~\omega_{c}(\omega_{c}+i\gamma)+
	\omega_{c}\sin^2{\alpha}(\omega-i\gamma)+\cos{\alpha}\sin{\alpha}(i\cos{\theta}
	(\omega^2+\gamma^2-\omega_{c}^2)-\sin{\theta}(\omega^2+\gamma^2+\omega_{c}^2))],
	\end{split}
	\end{equation}
	\begin{equation}
	\begin{split}
	&\mathbf{J_{dc}}=-q\gamma''\sin{\phi}\frac{(eE_0)^2 h e  V_{0}N}{8m^{3}s^{2}}\frac{1}{(\omega+i\gamma)D_{q\omega}\left[(\omega-i\gamma)^2-\omega_{c}^2\right](\gamma^2+\omega_{c}^2)}\begin{pmatrix}
	\gamma& -\omega_{c}\\
	\omega_{c}& \gamma
	\end{pmatrix}
	\begin{pmatrix}
	C_{x}&\\
	C_{y}
	\end{pmatrix}+c.c.
	\end{split}
	\end{equation}
	Introducing now Stokes parameters, after some algebra, we
reproduce Eqs.~\eqref{Jstart} and \eqref{R}  of the main text with
\begin{align}
	\label{a0}
	\m a_0 & = 2 \omega^2\omega_{\rm c} \left|\omega_{\rm c}^2 - (\omega-i \gamma)^2\right|^2
	\left[
	\begin{array}{c}
		-\omega_{\rm c} \\
		\gamma \\
	\end{array}
	\right]
	\\
	&+ s^2q^2 \left[
	\begin{array}{c}
		(\gamma^2+2\omega_{\rm c}^2)(\gamma^2+\omega^2)^2 +(\gamma^2-2\omega^2)\omega_{\rm c}^4 \\
		\gamma \omega_{\rm c} \left[ \gamma^4-\omega^4 +\omega_{\rm c}^4 + 2 \omega_{\rm c}^2(\gamma^2+2\omega^2)  \right]\\
	\end{array}
	\right],
	\nonumber
	\\
	\label{aL1}
	\m a_{\rm L1} & =s^2q^2 \gamma( \gamma^2+\omega^2+\omega_c^2)\left[
	\begin{array}{c}
		-\omega_{\rm c} (3 \gamma^2 +\omega^2-\omega_{\rm c}^2)\\
		\gamma ( \gamma^2 +\omega^2- 3\omega_{\rm c}^2) \\
	\end{array}
	\right],
	\\
	\label{aL2}
	\m a_{\rm L2} & =s^2q^2 \gamma( \gamma^2+\omega^2+\omega_c^2)\left[
	\begin{array}{c}
		\gamma( \gamma^2 +\omega^2-3\omega_{\rm c}^2)\\
		\omega_{\rm c} ( 3 \gamma^2 +\omega^2- \omega_{\rm c}^2) \\
	\end{array}
	\right],
	\\
	\label{aC}
	\m a_{\rm C} & =\omega (\omega^2+\gamma^2+\omega_{\rm c}^2)
	\left|\omega_{\rm c}^2 - (\omega-i \gamma)^2\right|^2
	\left[
	\begin{array}{c}
		\omega_{\rm c} \\
		-\gamma \\
	\end{array}
	\right]
	\\
	&+ s^2q^2 \omega( \gamma^2+\omega^2+\omega_c^2)\left[
	\begin{array}{c}
		-\omega_{\rm c} (3 \gamma^2 +\omega^2-\omega_{\rm c}^2)\\
		\gamma ( \gamma^2 +\omega^2- 3\omega_{\rm c}^2) \\
	\end{array}
	\right],
	\nonumber
\end{align}

\be D_{\omega q}=\omega (\omega+ i\gamma) -q^2s^2-\omega_{\rm c}^2
\frac{\omega}{\omega+ i\gamma}.  \label{D}\ee

\end{widetext}
\end{document}